# Measuring the Ability to Form a Product Line from Existing Products

Christian Berger, Holger Rendel, Bernhard Rumpe
RWTH Aachen University
Department of Software Engineering
Aachen, Germany
www.se-rwth.de

*Abstract*—A product line approach can save valuable resources by reusing artifacts. Especially for software artifacts, the reuse of existing components is highly desirable. In recent literature, the creation of software product lines is mainly proposed from a top-down point of view regarding features which are visible by customers. In practice, however, the design for a product line often arises from one or few existing products that descend from a very first product starting with copy-paste and evolving individually. In this contribution, we propose the theoretical basis to derive a set of metrics for evaluating similar software products in an objective manner. These metrics are used to evaluate the set of products' ability to form a product line.

## I. Introduction and Motivation

Recent literature regarding the creation of software product lines often proposes to use end-user visible characteristics of several products which are referred to as features [1], [2]. In most cases, common and variable attributes of a set of products are identified and a feature model is created [3], [4]. This is a high-level view and supports a top-down method for implementing product lines which bases on the assumption that the code structure can and will be organized according to the identified features. In practice, however, it often happens that a product line is only set up after one or even several similar product variants are implemented. Hence, it is inevitable to not only look at the desired features but also at the existing implementation to identify potential for reuse. Therefore, a bottom-up method is necessary to look especially at the implementation of these artifacts to identify commonalities and differences which either support or prevent the setup of a product line from a set of similar products.

In the following we present an approach which uses the software architecture and existing software artifacts of a set of similar products to evaluate their potential to form a product line. This approach bases on a set of metrics for measuring the so-called *product line-ability* of the considered set of products.

## II. Related Work

The authors of [5] and [6] describe the importance of product line scoping which is a top-down view on a product line. Reusable assets of existing products can be identified by a product vs. feature-matrix which can be implemented using different methodologies like generative programming [3]. The authors of [7] mention scoping as one aspect in number of steps when establishing a product line.

Metrics for evaluating product line architectures are discussed in several publications. The authors of [8] propose some metrics which are based on provided and required interfaces of components. However, these metrics are useful for object-oriented architectures only. A very formal specification of a product line architecture is given in [9] where parts of the architecture are treated as processes. In [10], some metrics are proposed to evaluate the quality of a product line which can only be applied for an existing product line with an already existing variability model.

The VEIA-project [11] also proposes very detailed metrics for product line architectures. Based on a function net and a feature model, these metrics measure the effort to integrate specific features into the product line. The use of function nets which define views on a so-called 150%-model of a product is also discussed in [12].

## III. Measuring the Product Line-Ability

In this section the theoretical basis for measuring the ability of a set of products to form a product line is outlined. Therefore, a set $\mathcal{P}$ containing $n$ similar products $p_1 \ldots p_n$ is evaluated. Herein, the term *similar* needs to be precisely refined by a set of metrics which evaluate the considered products in an objective manner.

### A. Specifying Similar Product Sets

As exemplarily shown in Fig. 1, a set $\mathcal{P}_3$ of three *similar* products $p_1$, $p_2$, and $p_3$ is shown for evaluating $\mathcal{P}_3$'s product line-ability. In this figure, three different classes named $C_1$, $C_2$, and $C_3$ of relations between two or more products are analyzed: $C_1$ describes the relation between two products, $C_2$ describes the reusability relation for commonly available parts for a specific product, and $C_3$ describes the reusability's benefit ratio for shareable parts for a specific product.

For evaluating a given set of similar products, each product is decomposed into $i = 1 \ldots n$ so called reasonable *atomic* pieces $c_{p_j,i}$ for a concrete product $p_j$ which is *self-contained* and *reusable*. We refer to these as *components* as defined in [2].

To perform a decomposition, all components $c_{p_j,i}$ must be identified and formally specified. Thus, we propose an annotated, directed graph $G_{p_j}$ for product $p_j$ which reflects the dependencies between all components $c_{p_j,i}$ which can for example be logical or communicative. The graph is defined as shown in Eq. (1).

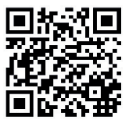

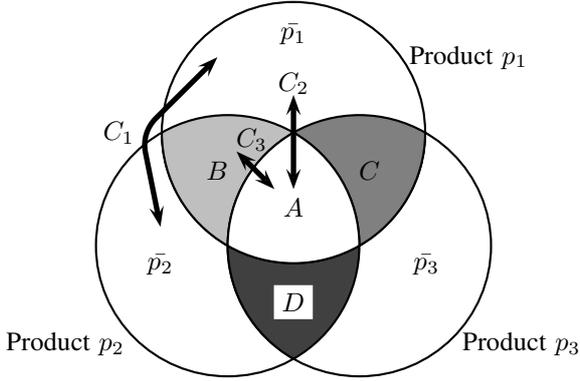

Fig. 1. Example for evaluating three similar products $p_1$, $p_2$, and $p_3$. The circles indicate the set of components for each product; $\bar{p}_2$ denotes the complementary set of components for product $p_2$ without the sets $B$, $A$, and $D$. $A$ denotes the set of components which are shared among all products; thus, all components in this intersection have at least a syntactically identical signature. $B$ denotes all components which are shared only by $p_1$ and $p_2$; $C$ and $D$ are calculated in an analog manner. $C_1$, $C_2$, and $C_3$ denote different classes of relations.

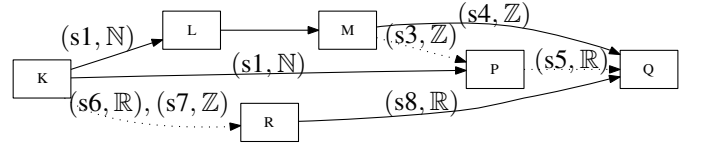

Fig. 2. An exemplary components' graph for six components $K$, $L$, $M$, $P$, $Q$, and $R$. The solid edges represent required communicative dependencies while the dotted edges represent optional ones. In this example, $K$ sends the same message to $L$ and $P$; $L$ sends an empty message to $M$ and thus simply calls it. Moreover, $K$ sends a message to $R$ consisting of two data fields.

$$\begin{aligned} G &:= (V, E) \\ V &:= id \\ E &:= V \times V \times \mathcal{P}(S) \times A \\ S &:= id \times \{\mathbb{N}, \mathbb{R}, [\![TYPE]\!], \dots\} \\ A &:= \{0; 1\} \end{aligned} \quad (1)$$

As shown in Eq. (1), the directed annotated graph $G$ consists of a set of ordered pairs of edges like $e_{1,2} = (c_1, c_2, (id, \mathbb{Z}), 0) \in E$. Each edge describes a formal dependency between the source component $c_1$ and its target component $c_2$ which reflects either a formal method call or a directed communication between component $c_1$ and $c_2$. In the former case, it describes the required signature $S$ in the target component for a successful method call, in the latter it defines a message which is sent from $c_1$ to $c_2$ containing the specified data in $S$. Components without any dependencies are so-called *isolated* components.

The set $A$ can be used to define *required* and *optional* components within a product; a value of 1 defines an optional while 0 defines a required dependency. The former defines a component which is inherently necessary to fulfill a product's so called *basis functionality*, while the latter adds further functionality like convenience functions; if unspecified, the edge is regarded as *required*.

In Fig. 2, a graphical representation of the aforementioned definition for the graph is shown for a product of six components is shown. Here, $\bar{p}_{r_1} = K, L, M, Q$ describes one path of *required* components, while $\bar{p}_{o_1} = K, P, Q$ describes one path of *optional* ones. For calculating the set $C_r$ of *required* and the set $C_o$ of *optional* components, recursive backtracking is used for all incident edges of an initially given set of components. Therefore, all product's components are initially added to the set $C_o$. Starting at a given required set of components $C_{start}$ from the considered product which can be for example some components for an actuator, all edges to adjacent components are analyzed. If a required edge is found it is added to $C_r$ which itself is analyzed recursively until all dependent required components are found. This set is finally subtracted from $C_o$. For example in Fig. 2 starting at $Q$, the following sets are calculated: $C_r = K, L, M, Q, R$ and $C_o = P$. The aforementioned algorithm does not identify isolated components because they do not contribute any reasonable data and thus, their relevance should be analyzed precisely.

### B. Metrics for Evaluating the Product Line-Ability

For evaluating the product line-ability of a set of $n$ similar products, the sets $C_{p_1,r} \dots C_{p_n,r}$ and $C_{p_1,o} \dots C_{p_n,o}$ with $\forall n : C_{p_n} \equiv C_{p_n,r} \cup C_{p_n,o}$ are calculated. Now, these sets can be evaluated according to Fig. 1. Therefore, different intersections between all sets are calculated which are used to evaluate different ratios and relations. For the sake of clarity, it is assumed that the denominator would not be 0 which means that two products do not share any components and thus, their comparison is not meaningful.

**Size of Commonality.**

$$SoC = \left| \bigcap_{i=1 \dots n} C_{p_i} \right| = \left| \bigcap_{i=1 \dots n} C_{p_i,r} \right| + \left| \bigcap_{i=1 \dots n} C_{p_i,o} \right|. \quad (2)$$

In Eq. (2), the *Size of Commonality* is shown which is calculated from set $A$ in Fig. 1 containing the number of identical components. It can be calculated by comparing the components' signatures: Two components are syntactical identical if they have the same signature. If *SoC* is 0, no commonly reusable components could be identified. This comparison is called *syntactical signature identity* which is at least *necessary* but not *sufficient*. Therefore, *semantic signature identity* for two components must additionally be ensured which can be for example be evaluated automatically by using the component's test suites in an entangled manner which have to ensure path coverage at least.

**Impact of Commonality.**

$$IoC = \frac{|\bigcap_{i=1...n} C_{p_i,r}|}{SoC}. \quad (3)$$

In Eq. (3), the *Impact of Commonality* is shown which relates *SoC* to all commonly shareable components. Obviously, the greater this ratio the more important are the commonly shareable components.

**Product-related Reusability.**

$$PrR_i = \frac{SoC}{|C_{p_i}|}. \quad (4)$$

The ratio in Eq. (4) describes the reusability of *SoC* for a specific product $p_i$: The greater this ratio the better is its reusability. This ratio is denoted by $C_2$ in Fig. 1.

**Impact of Product-related Reusability.**

$$IPrR_i = \frac{\left|\bigcap_{j=1...n} C_{p_j,r}\right|}{|C_{p_i,r}|}. \quad (5)$$

The ratio in Eq. (5) describes the impact of reusability of all commonly available components related to a specific product $p_i$ which is also denoted by $C_2$ in Fig. 1. Here, the smaller $1-IPrR_i$ for product $p_i$ the greater is the impact of all commonly shared components for this product.

**Reusability Benefit.**

$$RB_{i,j} = \frac{SoC}{|C_{p_i} \cap C_{p_j}|}. \quad (6)$$

In Eq. (6), the pairwisely calculated *Reusability Benefit* is shown which is denoted by $C_3$ in Fig. 1. For example, this ratio for $p_1$ and $p_2$ is calculated by $\frac{|A|}{|A|+|B|}$. The greatest quotient among all products describes the pair which shares the least commonly available components and vice versa.

**Relationship Ratio.**

$$RR_{i,j} = \frac{|C_{p_i} \cap C_{p_j}|}{|C_{p_i} \cup C_{p_j}|}. \quad (7)$$

In Eq. (7), the relationship between two products is calculated which is shown as $C_1$ in Fig. 1. Therefore, $A$ together with the number of components which are shareable between these two products only is related to the joined set of all remaining components of both products; the greater $RR_{i,j}$ between two products $p_i$ and $p_j$ the more similar are both products.

**Individualization Ratio.**

$$IR_i = \frac{\left|C_{p_i} \setminus \left(\bigcup_{k=1...n, k \neq i} C_{p_k}\right)\right|}{|C_{p_i}|}. \quad (8)$$

In Eq. (8), the product-related *Individualization Ratio* is calculated which describes the product's individualization related to the amount of components which are shared with at least one other product. The smaller this ratio the greater is this product's similarity with other products. In Fig. 1, this ratio is depicted by $IR_2 = \frac{|C_{p_2} \setminus (C_{p_1} \cup C_{p_3})|}{|C_{p_2}|}$ for product $p_2$.

## IV. APPLICABILITY OF THE METRICS

In the following, we apply the aforementioned metrics on a simplified example from the automotive domain for three different implementations of a door ECU. The first product as shown in Fig. 3 has only a lock/unlock functionality which locks the doors automatically at a specific vehicle's velocity. In Fig. 4, the product has no auto-lock function but power windows and a panic button to immediate closing in case of danger. Finally, in Fig. 5, a component exists to control window functions while opening or closing the hood of a convertible; this system also has an auto-lock function. All depicted signals have the same type.

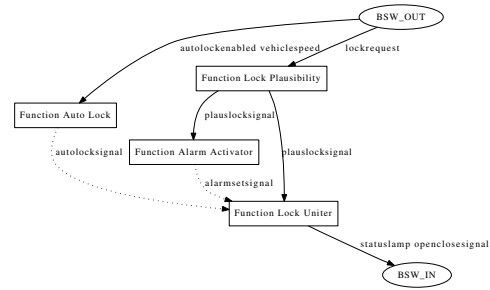

Fig. 3. Product $p_1$ "door ECU with auto-lock".

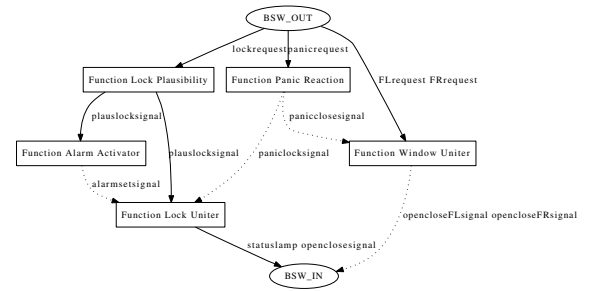

Fig. 4. Product $p_2$ "door ECU with power windows and a panic button".

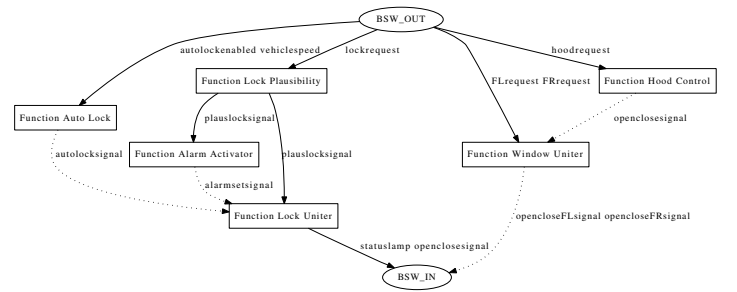

Fig. 5. Product $p_3$ "door ECU for convertibles".

To apply our metrics, we first have to determine the sets of products and their intersections. For the sake of clarity, the components are referred to by their abbreviation i.e. FLU for Function Lock Uniter. The *required* and *optional* components of the aforementioned products are shown in Tab. I.

TABLE I
REQUIRED AND OPTIONAL COMPONENTS

| product | required | optional |
|---|---|---|
| $p_1$ | FLP, FLU | FAL, FAA |
| $p_2$ | FLP, FLU | FAA, FPR, FWU |
| $p_3$ | FLP, FLU | FAL, FAA, FWU, FHC |

Now we are able to map these components to the corresponding sets as depicted by Fig. 1 and shown in Eq. (9).

$$\begin{aligned}
\bar{p}_1 &= \emptyset \quad (9)\\
\bar{p}_2 &= \{FPR, FLU_{p_2}, FWU_{p_2}\}\\
\bar{p}_3 &= \{FWU_{p_3}, FHC\}\\
A &= \{FLP, FAA\}\\
B &= \emptyset\\
C &= \{FAL, FLU_{p_{1,3}}\}\\
D &= \emptyset
\end{aligned}$$

The application of different metrics yields the results summarized in Tab. II.

TABLE II
RESULTS OF METRICS FOR EXAMPLE PRODUCTS

|  | all | $p_1$ | $p_2$ | $p_3$ | $p_{1,2}$ | $p_{1,3}$ | $p_{2,3}$ |
|---|---|---|---|---|---|---|---|
| number of components |  | 4 | 5 | 6 |  |  |  |
| SoC | 2 |  |  |  |  |  |  |
| IoC | 0.5 |  |  |  |  |  |  |
| PrR |  | 0.5 | 0.4 | 0.33 |  |  |  |
| IPrR |  | 0.33 | 0.33 | 0.33 |  |  |  |
| RB |  |  |  |  | 1 | 0.5 | 1 |
| RR |  |  |  |  | 0.29 | 0.67 | 0.22 |
| IR |  | 0 | 0.6 | 0.33 |  |  |  |

The results show that potential for reusability exists in general by *Size of Commonality*. *Impact of Commonality* has a value of 0.5 which means that the half of the common components are required. The product $p_1$ has to contribute to the product line because it has the highest *Product-related Reusability*. The *Impact of Product-related Reusability* is the same for all products and thus, no additional recommendation for a specific product to support the aforementioned ratio can be deduced. If *PrR* and *IPrR* for a specific product are small the product should not be part of the considered product line. The *Reusability Benefit* of $p_1$ and $p_3$ is the smallest because they share more than only the components of $A$. Besides, these products have also the highest *Relationship Ratio* which means they share the most common components if pairwisely compared and thus, they are suitable for a product line. The ratio *IR* indicates that $p_2$ has the highest amount of components which are independent from others. Hence, the product line should be created starting with the products $p_1$ and $p_3$; the product $p_2$ should be analyzed to identify potential for refactoring to improve its specific ratio of reusability.

## V. CONCLUSION

This paper outlined a collection of metrics for measuring the ability for a product line of a given set of products. First, the mathematical basis was discussed to summarize the necessary information without relying on a particular model which can be code excerpts, UML sequence charts, or AUTOSAR functional components for example. Using the mathematical model, several metrics are presented and their importance and benefit for a product line are considered. In a simplified example, these metrics are exemplarily used to show their application.

Currently, these metrics are applied at an industrial project from the automotive domain that should be transformed into a product line. Here, the goals are to evaluated the proposed metrics, identify necessary and sufficient commonalities as well as correlations, and to estimate a set of values which recommends the creating of a product line. Another goal is to have a closer look on the models which describe software artifacts and their transformation into a suitable representation which we use as basis for the metrics.

REFERENCES

[1] P. Clements and L. Northrop, *Software Product Lines: Practices and Patterns*. Addison-Wesley, 2002.
[2] K. Pohl, G. Böckle, and F. Linden, *Software Product Line Engineering: Foundations, Principles, and Techniques*. Springer, 2005.
[3] K. Czarnecki and U. W. Eisenecker, *Generative Programming: Methods, Tools, and Applications*. Addison-Wesley, 2000.
[4] K. Kang, S. Cohen, J. Hess, W. Nowak, and S. Peterson, "Feature-oriented domain analysis (foda) feasibility study," Technical Report CMU/SEI-90-TR-21, Software Engineering Institute - Carnegie Mellon University, Tech. Rep., 1990.
[5] J. Bosch, *Design and Use of Software Architectures: Adopting and Evolving a Product-Line Approach*. New York, NY, USA: ACM Press/Addison-Wesley Publishing Co., 2000.
[6] I. John and M. Eisenbarth, "A decade of scoping - a survey," in *Proceedings of the 13th International Software Product Line Conference*, 2009.
[7] P. C. Clements, L. G. Jones, J. D. McGregor, and L. M. Northrop, "Getting there from here: a roadmap for software product line adoption," *Commun. ACM*, vol. 49, no. 12, pp. 33–36, 2006.
[8] E. Dincel, N. Medvidovic, and A. v. d. Hoek, "Measuring product line architectures," in *PFE '01: Revised Papers from the 4th International Workshop on Software Product-Family Engineering*. London, UK: Springer-Verlag, 2002, pp. 346–352.
[9] A. Gruler, M. Leucker, and K. Scheidemann, "Calculating and modeling common parts of software product lines," *Software Product Line Conference, International*, vol. 0, pp. 203–212, 2008.
[10] T. Zhang, L. Deng, J. Wu, Q. Zhou, and C. Ma, "Some metrics for accessing quality of product line architecture," *Computer Science and Software Engineering, International Conference on*, vol. 2, pp. 500–503, 2008.
[11] S. Mann and G. Rock, "Dealing with variability in architecture descriptions to support automotive product lines: Specification and analysis methods," in *Proceedings embedded world Conference 2009*. Nürnberg, Deutschland: WEKA Fachmedien, Mar. 3-5, 2009.
[12] H. Grönniger, J. Hartmann, H. Krahn, S. Kriebel, L. Rothhardt, and B. Rumpe, "Modelling automotive function nets with views for features, variants, and modes," in *Proceedings of ERTS '08*, 2008.